\documentclass[usenatbib]{mn2e}
\usepackage[dvips]{graphicx}
\usepackage{amssymb}
\usepackage{natbib}

\begin{document}

  \title[An observer's view of simulated galaxies]{An observer's view of simulated
    galaxies:  disc-to-total ratios, bars, and (pseudo-)bulges}

\author[Scannapieco et al.]{Cecilia Scannapieco$^{1,2}$,
Dimitri A. Gadotti$^{2,3}$, Patrik Jonsson$^{4}$
and  Simon D.M. White$^{2}$
\\
$^1$ Astrophysical Institute Potsdam, An der Sternwarte 16, D-14482, Potsdam, Germany\\
$^2$ Max-Planck Institute for Astrophysics, Karl-Schwarzschild Str. 1, D-85748, Garching, Germany\\
$^3$ European Southern Observatory, Casilla 19001, Santiago 19, Chile\\
$^4$ Institute for Theory and Computation, Harvard-Smithsonian Center for
Astrophysics, 60 Garden St., MS-51, Cambridge, MA 02138, USA\\
}

   \maketitle

   \begin{abstract}

We use cosmological hydrodynamical simulations of the formation
of Milky Way-mass galaxies to study the relative
importance of the main stellar components, i.e., discs, bulges, and bars, at redshift zero.
The main aim of this work is to understand
if estimates of the structural parameters of these components
determined from kinematics (as is usually
done in simulations) agree well with those obtained using
a photometric bulge/disc/bar decomposition (as done in observations).  
To perform such a comparison,
we have produced synthetic observations of the simulation outputs
with the Monte-Carlo radiative transfer code {\sc sunrise} 
and used the {\sc budda} code to make 2D photometric decompositions of the
resulting images (in the $i$ and $g$ bands). 
We find that the kinematic disc-to-total ratio (D/T)  estimates are systematically and significantly
lower than the photometric ones.  While the maximum D/T ratios obtained
with the former method are of the order of $0.2$, they
are typically  $>0.4$, and can be as high as $0.7$, according to the
latter. 
The photometric decomposition shows that many of the simulated galaxies have bars, with Bar/T ratios in the range $0.2-0.4$, and that bulges have in all cases low S\'ersic indices, 
resembling observed pseudo-bulges instead of
classical ones.
Simulated discs, bulges and bars  generally have similar $g-i$ colours, which are in the blue tail of the distribution of observed colours. 
This is not due to the presence of young stars, but rather to low metallicities and poor gas content in the simulated galaxies, which makes dust extinction low.
Photometric decompositions thus match the component ratios usually
quoted for spiral galaxies better than kinematic decompositions, but
the shift is insufficient to make the simulations consistent with
observed late-type systems.
   \end{abstract}


\section{Introduction}

In the local universe, a significant fraction of the stellar mass is
observed to be in discs ($\sim 60$ per cent, \citealt{Driver07}; \citealt{Weinzirl} -- see also \citealt{Gadotti09}).
This is difficult to reconcile with simulations of galaxy
formation in a $\Lambda$CDM universe, where hierarchical assembly tends to
produce systems with a large fraction of their stellar mass in a
bulge. In recent years,  the inclusion of efficient treatments
of supernova (SN) feedback, together with improved
numerical resolution,
have produced simulated disc galaxies more similar to real spirals
(e.g. \citealt{Brook04}; \citealt{Gov04}; \citealp{S08,S09}  and references therein),
although although in most cases these are still
dominated by old, centrally concentrated spheroids. 
A limitation of these studies is the fact that comparison between
simulations and observations is often rather crude, and the methods
applied to derive structural parameters of simulated and observed galaxies
are often very different. As a consequence, the results are not
directly comparable, and  it is hard to decide
how close or how far from reality the simulated galaxies are. This is
a serious problem since, in order to improve galaxy formation models,
we need to know where we fail. Some recent studies have
tackled this problem by analysing the simulated data 
to obtain 'observables' more directly comparable to observational
results (e.g. \citealt{Gov09}; \citealt{Gov10}).

The main goal of this Letter is to perform a meaningful comparison
between observations and simulations. To this end, we use
a method we refer to as {\it photometric decomposition}, which consists
of first producing
synthetic images of simulated galaxies, and then
performing 2D bulge/disc/bar decompositions to obtain their structural
parameters and  colours.
In this
way, we are able to mimic real observations and 
to directly compare, e.g., the derived 
scale-lengths, S\'ersic indices, disc-to-total ratios, and colours  with
observational results.
The simulations we use are those studied in  \citet[hereafter S09]{S09},
which are cosmological hydrodynamical
simulations of the formation of Milky Way-mass galaxies in a  $\Lambda$CDM universe.
Using a {\it kinematic}
decomposition of the stars in the simulated galaxies, S09 found that 4/8 systems
have significant disc components in rotational support, but 
the maximum disc-to-total (stellar mass) ratios they obtained is of the order
of $0.2$.

Here we analyse these simulations using 
the photometric decomposition technique which \citet{Gadotti09} applied to
real galaxies to obtain structural parameters for discs, bars, 
classical bulges and pseudo-bulges. Gadotti's sample is
particularly
useful for a comparison, since it comprises galaxies
with masses similar to that of the Milky Way (as in the simulations).

This Letter is organized as follows. In Section~\ref{methodology}
we briefly describe the main characteristics of the simulations used
in this study, as well as the methods to perform the kinematic
and photometric decompositions. 
In Section~\ref{results} we show and discuss the outcome of the 
different techniques employed and compare  
results from photometric decompositions in simulated and real galaxies.
Section~\ref{conclu}  summarizes our results and conclusions.

\section{Methodology}\label{methodology}

\subsection{Simulation setup}

For this study, we use the redshift $z=0$ outputs of the simulations presented
in S09, which correspond
to eight galaxies with $z=0$ masses similar  to the Milky Way, assembled in the context of a
$\Lambda$CDM cosmogony ($\Omega_\Lambda=0.75$, $\Omega_{\rm m}=0.25$,  $\sigma_8=0.9$ 
and  $H_0=73$ km s$^{-1}$ Mpc$^{-1}$). 
The initial conditions (ICs)   are 
based on those generated for the Aquarius Project \citep{Springel08} of the Virgo
Consortium. To these dark-matter only ICs, we 
 added baryons assuming $\Omega_{\rm b}=0.04$. 
Target haloes were selected 
to satisfy a mild isolation  criterion at $z=0$ (no neighbour exceeding half
of their mass within $1.4$ Mpc). 
All simulations have similar mass resolution, with dark matter and gas
particle masses of the order of $2\times 10^6$ and $3\times 10^5$ M$_\odot$, respectively
(see  table~$1$ of S09 for details). We have used  the same  gravitational 
softening for dark matter, gas and star particles, which
varies in the range $0.7-1.4$ kpc for the different simulations.
The halos have final 
masses in the range $7-16$ $\times 10^{11}$ M$_\odot$, span a wide
range in spin parameters (between $0.008$ and $0.049$), and have
different merger and accretion histories. As a result, the galaxies present a variety
of $z=0$ morphologies (section~3 in S09).

The simulations have been run with the Tree-PM smoothed particle hydrodynamics (SPH) 
code {\sc gadget-3} \citep{Springel08}, 
with the additional implementation of star formation and feedback as
described in \citet{S05,S06}. Our model
includes stochastic star formation,  metal-dependent cooling, chemical enrichment and feedback from Type II 
and Type Ia SNe, with a multi-phase model for the gas component which avoids
excessive cooling found in standard
formulations of SPH, and allows winds to be generated
naturally without additional {\it ad hoc} assumptions. 
We note that our implementation of  star formation and feedback is
different from that of \citet{SH03}, although we do use their
treatment of UV background. 
More details on the simulation code, ICs
and input parameters can be found in S09.

The $i$-band total magnitude of the simulated galaxies is in the range from $-21$ to $-22$, which is similar to that of galaxies with stellar mass similar to that of the Milky Way, in the SDSS database \citep{KauHecWhi03}. Analogous results were found in the simulations by \citet{Gov04} and \citet{2003ApJ...591..499A}.

\begin{figure}
\begin{center}
{\includegraphics[width=30mm]{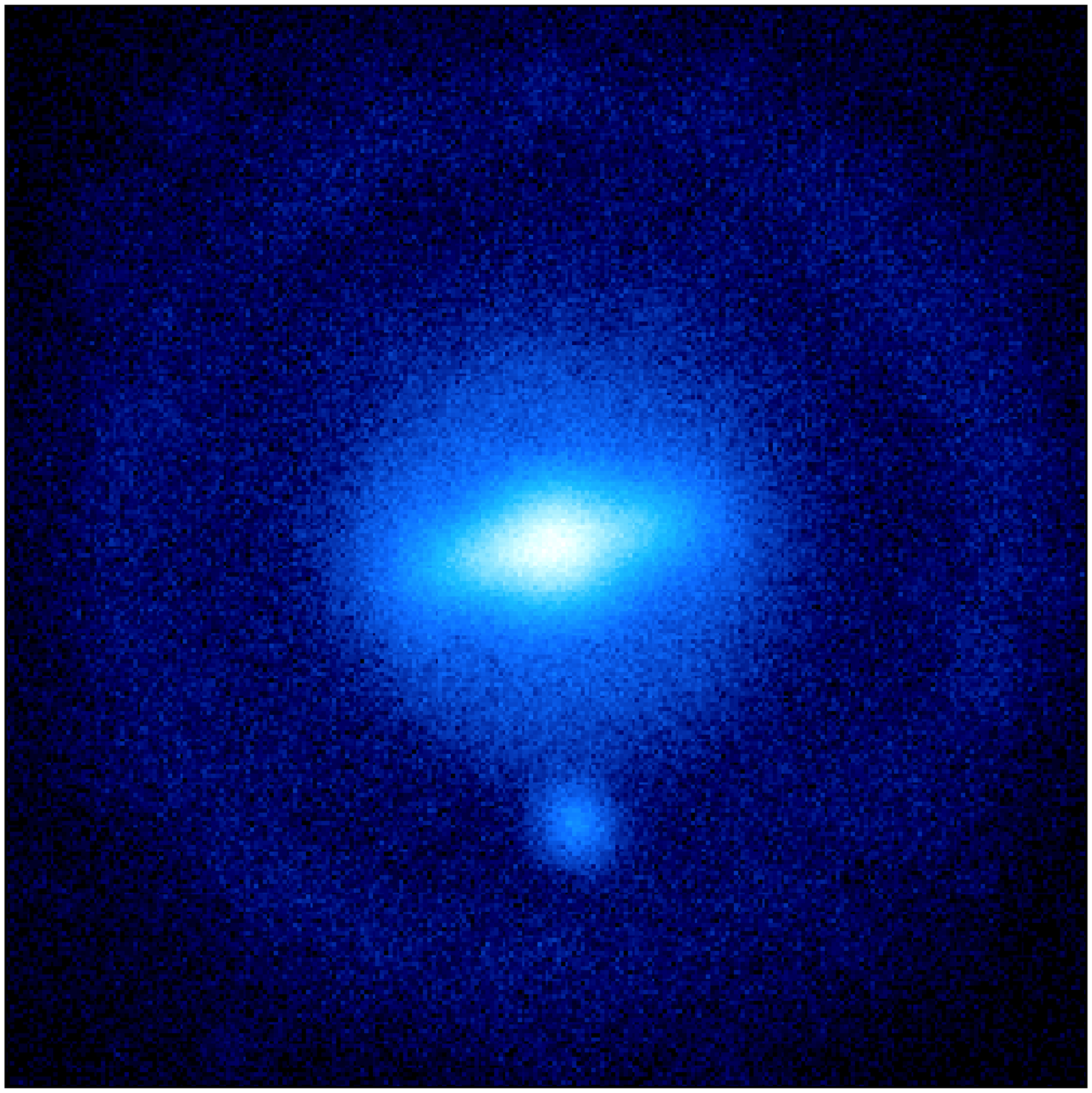}\includegraphics[width=30mm]{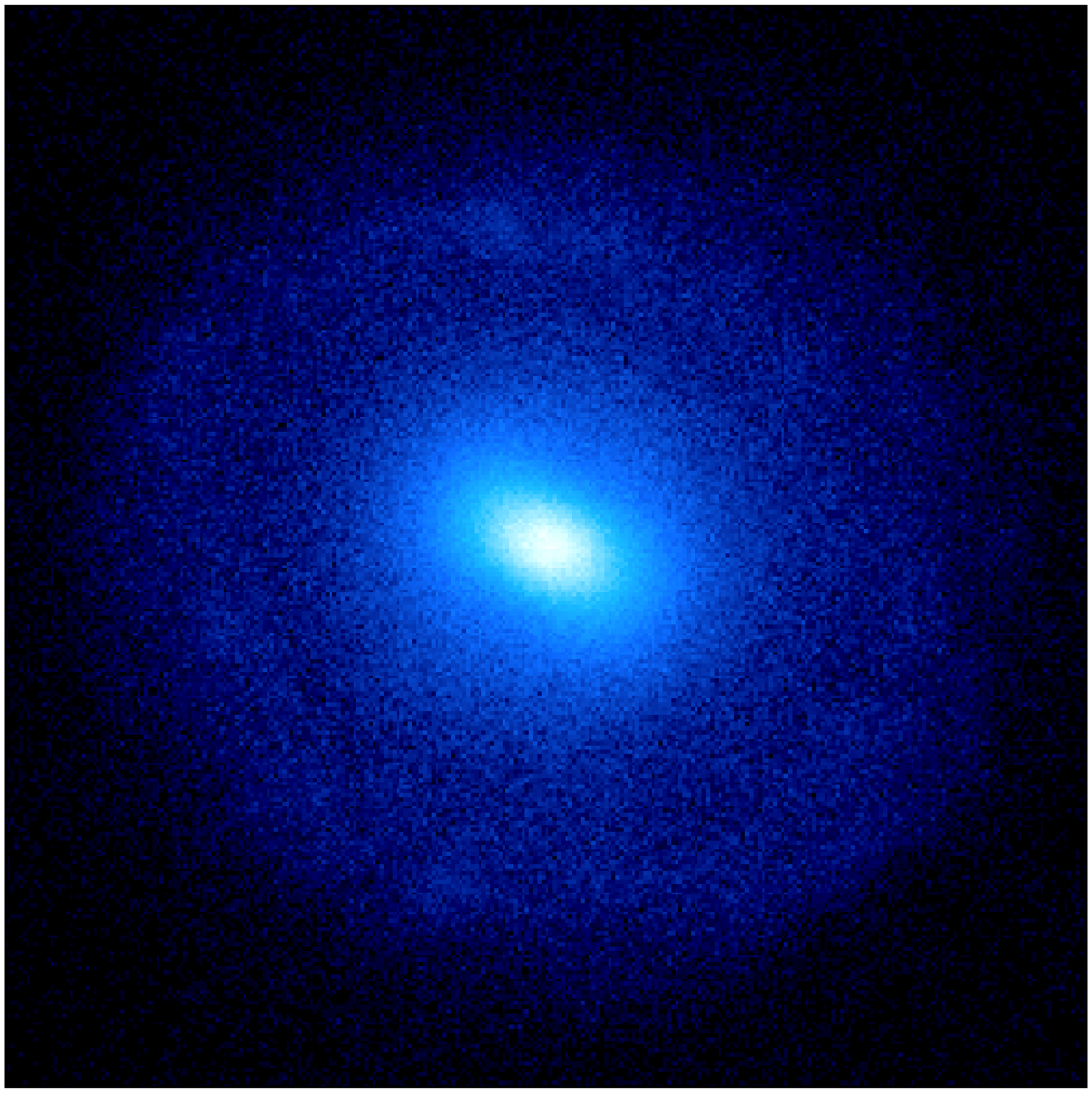}}\\

{\includegraphics[width=30mm]{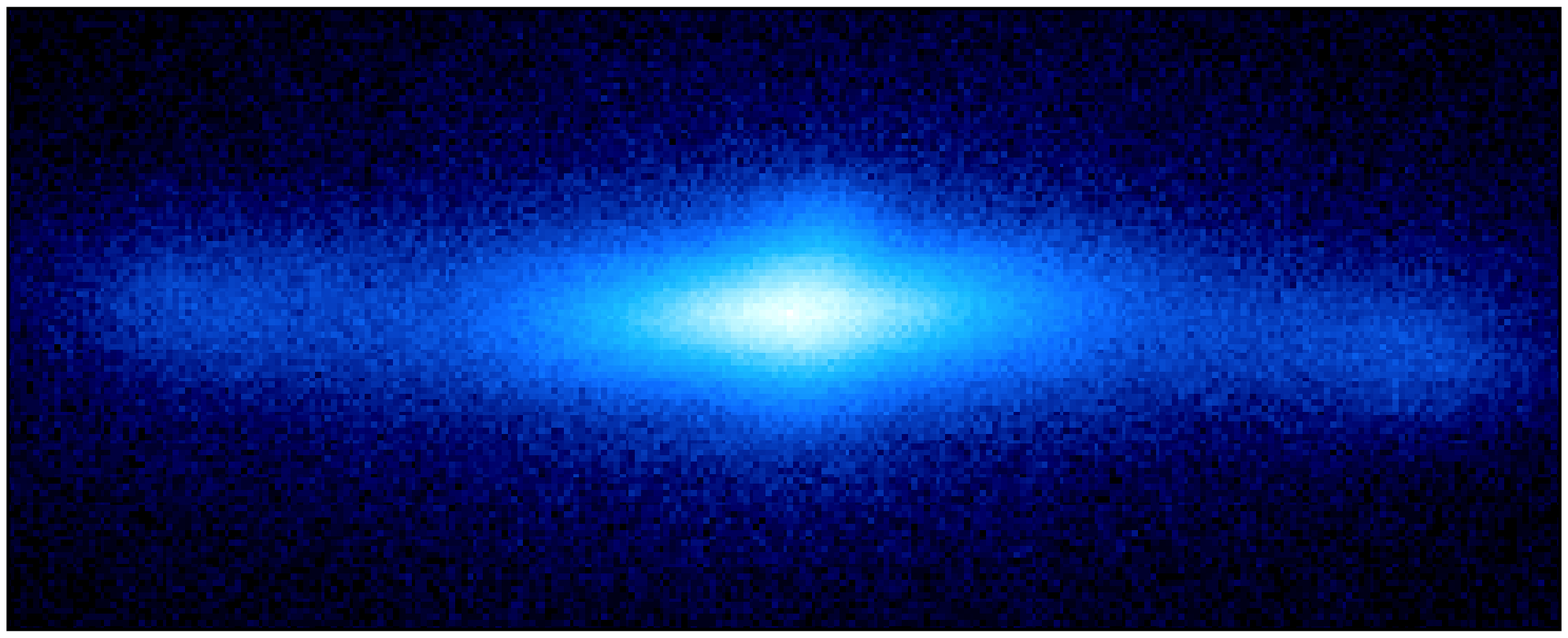}\includegraphics[width=30mm]{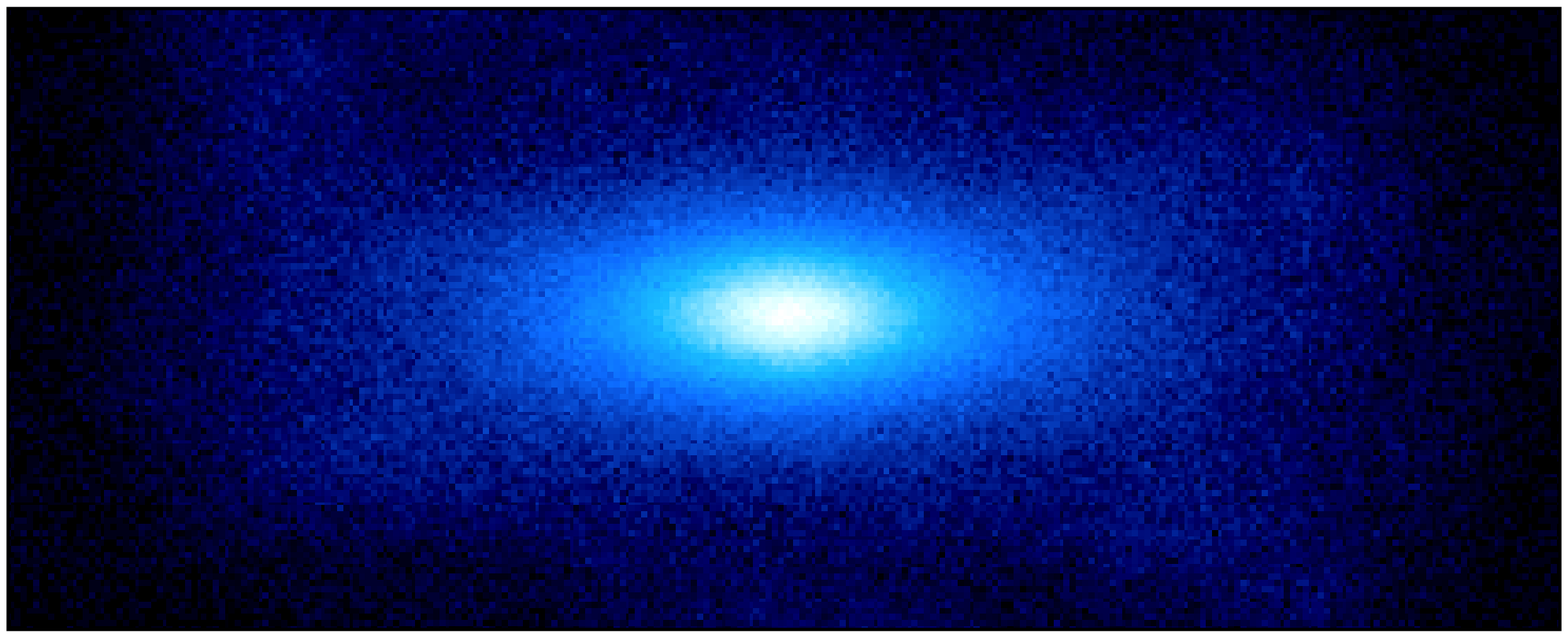}}\\
\caption{\small Face-on and edge-on SDSS $i$-band surface brightness images for Aq-C-5 (left-hand panels)
and Aq-G-5 (right-hand panels), at $z=0$. The images are $50$ kpc across, and the
edge-on images have a vertical depth of $20$ kpc.}
\label{fig1}
\end{center}
\end{figure}

\subsection{Photometric and kinematic  decompositions}

The {\it photometric} decomposition (PD) is done in two steps.
First, we use the Monte-Carlo radiative-transfer code {\sc sunrise}
\citep{Patrik, jgc09} to generate face-on, 
synthetic images of the simulated galaxies in the $g$ and $i$ SDSS bands
(examples of images are shown in Fig.~\ref{fig1}). 
{\sc sunrise} calculates the appearance of the simulated
galaxy, from far-ultraviolet to submillimeter wavelengths, by tracing
emitted radiation from the stellar particles through the dusty ISM of
the galaxy, assuming that the density of dust grains traces the
density of metals in the gas phase. The emission from dust grains is
calculated for every location in the galaxy based on the intensity of
radiation heating the dust grains, and is iterated to equilibrium. The
final output is the spatially resolved emerging radiation from the
simulated galaxy from a number of viewing angles, which can be
directly compared to real observations. For more details about the
radiation transfer calculation, the reader is referred to the above references.

\begin{figure}
\begin{center}
{\includegraphics[width=50mm]{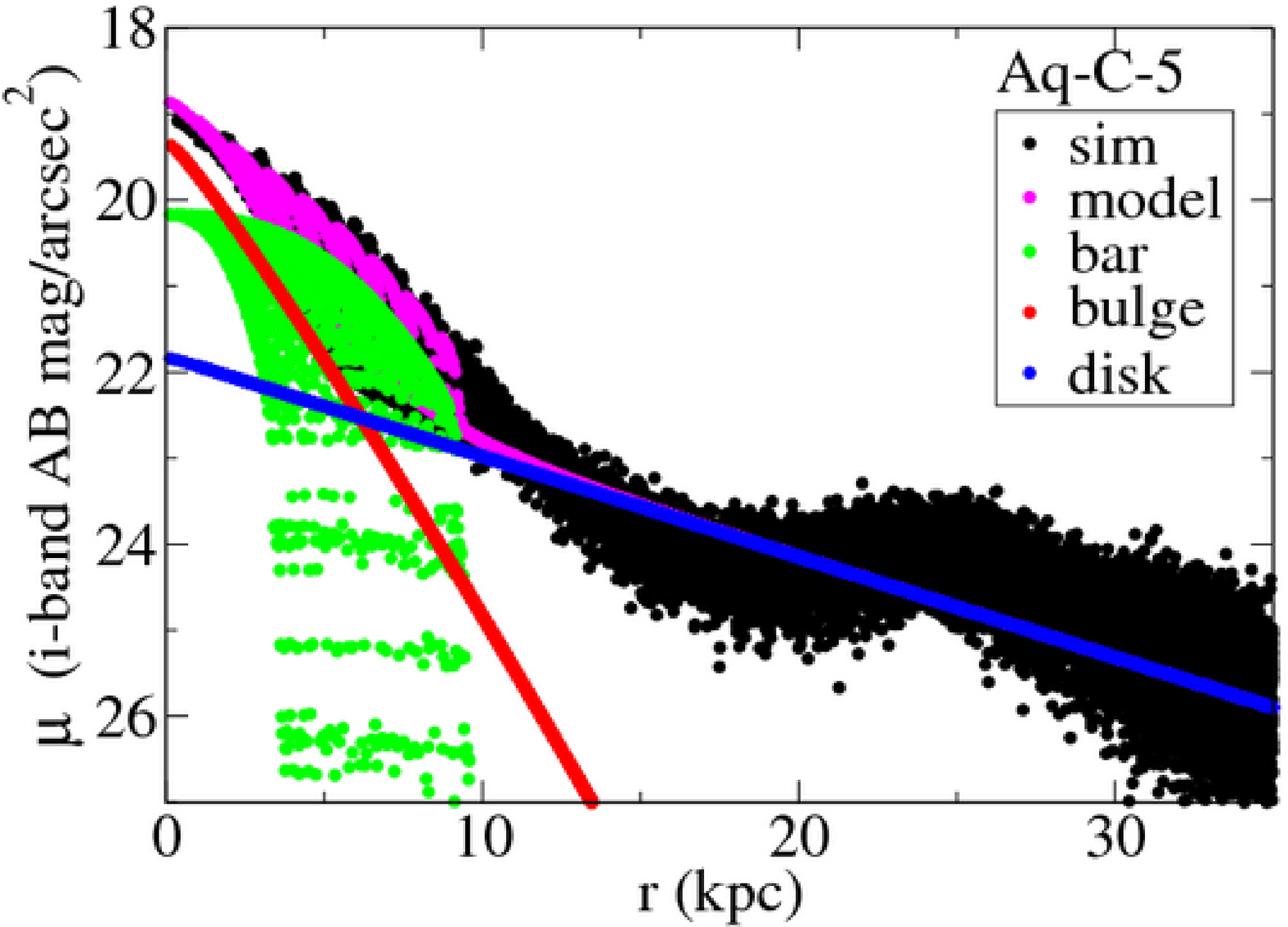}
\includegraphics[width=50mm]{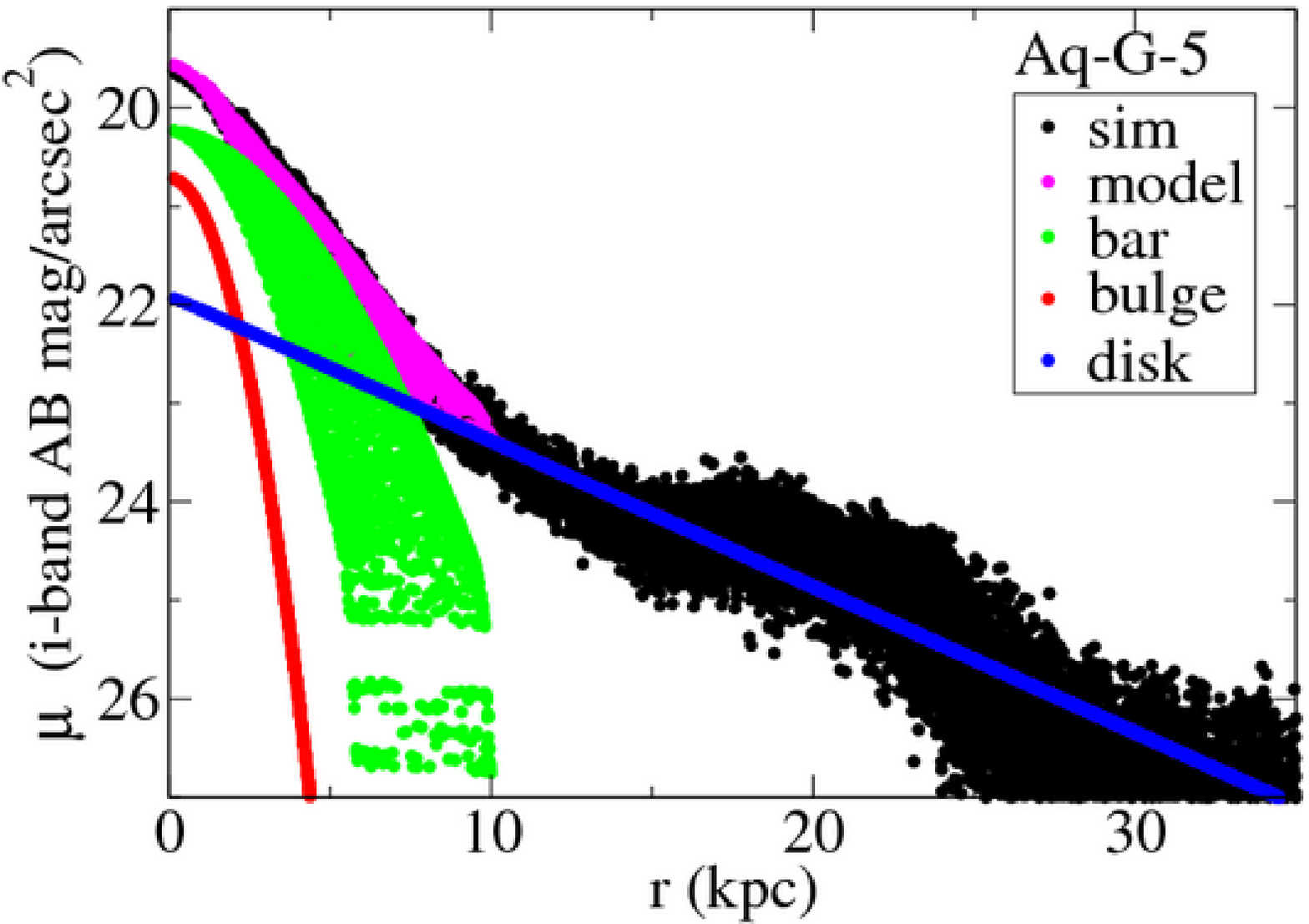}}
\caption{\small Results of the image decomposition (SDSS $i$ band) for Aq-C-5
(upper panel) and Aq-G-5 (lower panel):
surface brightness profiles of the simulated
galaxies and the contribution of the different model components, as indicated. 
Each point
corresponds to a single pixel (see \citealt{Gadotti08} for details).
}
\label{fig2}
\end{center}
\end{figure}

To mimic the effects of a real observation, the {\sc sunrise} images were 
convolved with a circular Gaussian function with FWHM of 1.5 pix, 
which corresponds to about 375 pc, or 0.75 arcsec, for a galaxy at
 a redshift of 0.03 and a plate scale of 0.5 arcsec per pix. With 
the same goal, we have multiplied these images by a factor such 
that their central pixels have generally several thousand ADU, and 
also added a background pedestal of 200 ADU. As a second step, we perform a 
2D bulge/disc/bar decomposition of the images
using the {\sc budda} code \citep{Gadotti08}, and
estimate the disc, bulge, and bar-to-total ratios, as well
as the disc scale-length, bulge effective radius, and bulge
S\'ersic index. 
By combining results from the $g$ and $i$
bands, 
we also obtain the integrated $g-i$ colours of all components. The inclusion of bars
is important since, as shown in e.g. \citet{Gadotti08},
neglecting them (when present) in the modelling of galaxies
can lead to large uncertainties in the estimation of bulge parameters, and a 
systematic overestimation of the bulge-to-total (B/T) ratio.
In Table~\ref{decomp_results}, we show the results of the PD, and in
Fig.~\ref{fig2}  we show the decomposition results
for the two galaxies shown in Fig.~\ref{fig1}. There are two cases, Aq-F-5  and Aq-H-5, for which a reliable decomposition
was not possible, as these systems present unusual structural components. If these were real observations, these
two galaxies would probably be excluded from the discussion. For completeness, we include
these results, but they are always highlighted to remind the reader that they might not be reliable.

The {\it kinematic} decomposition (KD)
assigns stars in the simulated galaxies to either a disc or a spheroidal
component (see S09 for details).
The decomposition is based on the distribution of $\epsilon\equiv j_{\rm z}/j_{\rm circ}$,
where $j_{\rm z}$  is the angular momentum
of each star perpendicular to the disc plane 
and $j_{\rm circ}$ is the angular momentum for a circular
orbit at the same radius. We combine the distribution of $\epsilon$ with
the radii of stars to define a disc component which does not suffer
from contamination of  spheroid stars in the inner regions (see Figs. $3$ and $4$ of S09).
The spheroidal
component is defined by stars that have not been  tagged as disc stars and, as
a result,
it includes not only bulge and stellar halo stars, but
also bar stars, when bars are present 
(visual inspection of the simulated galaxies indicates the presence of clear bars
in about half the simulated galaxies).
Once disc and spheroid stars are identified, we estimate
the (mass-weighted) disc-to-total (D/T$^{\rm k}$, shown in Table~\ref{decomp_results}) 
and spheroid-to-total  (1-D/T$^{\rm k}$) ratios, 
and the half-mass radii for both components. 
In the case of the disc, to be able to meaningfully compare
the spatial scale with that obtained with the PD, we convert
the half-mass radii into disc scale lengths, assuming an
exponential surface mass density profile. The relation is such
that the half-mass radius is $1.68$ times the {\it scale-length}.
Although, in some cases, the simulated disc profiles are not purely exponential,
in particular because of the presence of outer rings,
the assumption of an exponential profile is not expected to introduce
errors, since the PD anyway assumes exponential profiles and, moreover,
outer rings are also present in real galaxies.

\section{Results}\label{results}

\begin{table*} 
\centering
\begin{minipage}{177mm}
\caption{Results from the kinematic and photometric 
decompositions of the eight simulated galaxies. 
}
\small
\label{decomp_results}
\begin{center}
\begin{tabular}{lccccccccccccc}
\hline
Galaxy  & D/T$^{\rm k}$ & D/T$^{\rm p}$ & B/T$^{\rm p}$ &
Bar/T$^{\rm p}$ & $(g-i)_{\rm d}$ &$(g-i)_{\rm b}$ &$(g-i)_{\rm bar}$ & $r_{\rm d}^{\rm k}$ & $r_{\rm d}^{\rm p}$ & $r_{\rm s}^{\rm k}$ & $r_{\rm eff}^{\rm p}$ & $n$ \\
(1) & (2) & (3) &(4) & (5) & (6) &(7) & (8) & (9) &(10) & (11) & (12)&(13)\\
\hline
Aq-A-5   & 0.06  & 0.32 & 0.45 & 0.23 & 0.27
& 0.51 & 0.79 & 12.7 & 12.8 & 2.9 & 2.8  & 1.09 \\
Aq-B-5   & 0.09  & 0.42 & 0.58 & -   & 0.43
& 0.55 & -  & 14.4 & 15.7  & 3.9  & 2.6 & 0.97  \\
Aq-C-5   & 0.21  & 0.49 & 0.28 & 0.23 & 0.51
& 0.43 & 0.78 & 7.3 & 10.7 & 4.0 & 3.7  & 0.91 \\
Aq-D-5   & 0.20  & 0.68 & 0.32 & -    & 0.46
& 0.65 & -  & 6.6 & 7.6 & 4.0  & 2.5  & 0.64 \\
Aq-E-5   & 0.14  & 0.40 & 0.17 & 0.43 & 0.49
& 0.62 & 0.51 & 7.7 & 6.9 & 3.2 & 2.8  & 0.28 \\
Aq-F-5   & -     & (0.44) & (0.56) &  - & (1.02) & (0.59)
& (0.51) &- & (13.4) & 6.1 &  (4.3)  & (1.02) \\
Aq-G-5   & 0.23  & 0.60 & 0.06 & 0.34 & 0.39
& 0.06 & 0.69 & 6.5 & 8.4  & 3.2 & 1.7 & 0.50 \\
Aq-H-5   & 0.04  & (0.05) & (0.95) &  -  & (1.34) &  (-1.12)
& (0.69)  & 6.3 & (5.7) & 4.2  & (6.1) & (1.34) \\
\hline
\end{tabular}
\end{center}
Column (1) gives the designation of the simulated galaxy. Column (2)  gives the disc-to-total  estimated from the KD. Columns (3)-(5) show the disc, bulge and bar-to-total ratios obtained using the (SDSS $i$-band) photometric  decomposition.
Columns (6)-(8) give the $(g-i)$ colours of the disc, bulge and bar
components, respectively. 
We give in columns (9) and (10) the disc scale-lengths (in kpc)
obtained with
the kinematic and photometric ($i-$band) decompositions, respectively.
The kinematic half-mass radii for the spheroidal component
and photometric ($i-$band) bulge effective radii are shown in columns (11) and
(12) (both in kpc), and column (13) gives the
photometric ($i-$band) bulge S\'ersic index ($n$).
Results for Aq-F-5 and Aq-H-5 obtained from the PD are given in parenthesis
to remind the reader that these might not be reliable. 
\end{minipage}
\normalsize
\end{table*}

\subsection{Disc, bulge and bar to total ratios}

\begin{figure}
\begin{center}
{\includegraphics[width=85mm]{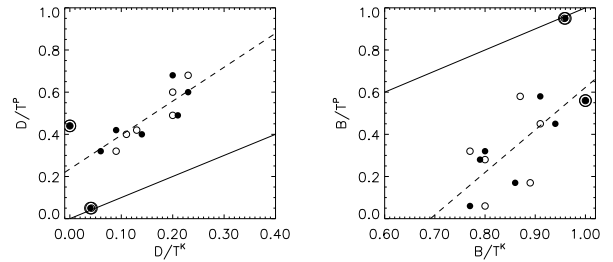} } 
\end{center}
\caption{\small Photometric versus kinematic results for the D/T
  and B/T  ratios. In the case of the KD, we include mass-weighted 
(filled symbols) and luminosity-weighted  (open symbols) estimates.
The solid lines depict the one-to-one correspondence, and the dashed lines
a linear fit to the data, excluding Aq-F-5 and Aq-H-5 (encircled symbols) 
for which the results of the PD might not be reliable.
The observed D/T in local late-type galaxies, with stellar masses similar to that of the Milky Way, ranges from about 0.7 to 1, while for B/T the corresponding value ranges from 0 to 0.2 \citep{Gadotti09}.}
\label{DTBT}
\end{figure}

The main result of this study is that the disc-to-total (D/T) ratios obtained
from the kinematic and photometric decompositions differ significantly 
(Table~\ref{decomp_results} and Fig.~\ref{DTBT}). 
In most cases, the  photometric D/T estimates are significantly higher than the
kinematic ones. 
According to the
KD, galaxies with significant disc
components typically have D/T$\sim 0.2$ while the PD yields
estimates of $0.4-0.7$.
Moreover, for those galaxies with kinematically not very massive
discs (D/T$\lesssim 0.09$), the photometric estimates
can be as high as $\sim 0.4$. 

Although the KD estimates concern mass, while those from PD concern luminosity,
this does not explain in this case such a discrepancy. As we will discuss further below,  
the mean stellar ages of the spheroid and disc components in the simulations are similar
{\em and} they are both sufficiently old, meaning that differences in mass-to-light ratio between
the
different components are small.
In fact, we find no significant difference when 
luminosity-weighted estimates (in the SDSS $i$ band),
calculated  using
the ages and metallicities of stars as inputs for the \citet{BruzualCharlot}
population
synthesis models (for a Salpeter initial mass function as assumed in the simulations), are used
(cf. open symbols in Fig.~\ref{DTBT}). 
Note that, in situations where disc and bulges have very different ages, this effect might however
become important (\citealt{Abadi}).
These results clearly indicate the
importance of comparing simulations with observations in
an appropriate manner. According to our results, estimates for
D/T ratios obtained from a kinematic decomposition
can be much lower than those obtained from 
photometry by observational methods.

A possible  source of the discrepancy between the D/T ratios obtained
with the photometric and kinematic methods 
is that the former approach assumes an exponential
profile for the discs, extending to the very centre.
On the contrary, in the KD, disc stars
are not present in the inner $~2$ kpc (see
figure $4$ in S09), since these central regions are purely
dominated by velocity dispersion. In fact,  
the contribution of the inner $2$ kpc 
to the total mass/luminosity for an exponential profile varies
between $15$ and $29$ per cent, for $r_{\rm d}=6$ and $14$ kpc (the
range of scale lengths we find for our simulations). In other words, if
we were to take kinematically identified disc particles and to fit
an exponential profile, we would get higher kinematic D/T estimates.
We note, however, that this effect alone can not fully
account for the differences in the D/T ratios between the two methods.

The differences in the D/T ratios  can not be
attributed to the particular band used in the analysis either.
In fact, we have performed the PD for the $g$-band as well, and the
results are similar to those obtained for the $i$-band. In particular,
the B/T ratios and S\'ersic indices change 
by less than $10$ per cent (with the exception of Aq-G-5 where the
$g$-band estimate is $45$ per cent higher than the $i$-band one), 
and no systematics
are detected. As for the D/T ratios, they are systematically
$\sim 5$ per cent  larger in the $g$-band (except for Aq-A-5, with
a $20$ per cent change).
On the contrary, the Bar/T ratios are typically 20 per cent lower
in the $g$-band. These changes are
somewhat expected, considering that the youngest stars are in the discs,
and bars are usually populated by old stars. It should also be noted that
dust effects are negligible in the simulated images, due to their low metal and
gas content. In fact, the typical face-on optical depth at the $V$ band in the
simulated galaxies is $\tau_V\approx 0.05-0.1$, which does not produce
noticeable effects in bulge/disc decompositions \citep[see][]{Gad10}.

Another source of discrepancy between the D/T ratios obtained with
the two methods is that a significant fraction ($\sim 40$ per cent) of the stars 
outside the inner regions ($r>5$ kpc) does not have disc-like kinematics; 
consequently they are not counted as disc particles in the KD, but
they do contribute to the disc in the PD.
The increase in the photometric D/T ratio due to this effect
is, nevertheless, not expected to be very high, since the light
from the disc
region is always dominated by young stars on near-circular orbits.

Another result of the photometric analysis is the
detection of bars, which, when present, can be
quite prominent, with Bar/T between $0.23$ and $0.43$ (Table~\ref{decomp_results}), 
values typical of observed strong bars \citep{Gadotti08}.
This also means that, even for moderate D/T values, the bulges do not necessarily
dominate over discs (Table~\ref{decomp_results}). In fact, all galaxies with bars have  D/T$>$B/T.
One interesting finding is that Aq-G-5 has a quite low B/T: only $6$ per cent of the
stellar luminosity is associated to the bulge. In this galaxy, the disc and bar contributions 
to the total luminosity are $60$ and $34$ per cent, respectively.
We also find cases of bulge-dominated galaxies
(the highest B/T we find are of the order of $0.6$), such
as Aq-A-5 and Aq-B-5 (as well as Aq-F-5 and Aq-H-5). Among these, some have bars and some do not.

\subsection{$g-i$ Colours, spatial scales and S\'ersic indices}

The $g-i$ integrated colours of the different components (bulge, disc and bar) 
were obtained directly from the {\sc budda} models corresponding to the $g$ and
 $i$ band images. We find that, generally, the three components have similar 
$g-i$ colours, which differ typically by only $0.1-0.2$ mag 
(Table \ref{decomp_results}). 
Furthermore, all components show $g-i$ colours which are in the blue tail of
the corresponding distribution for real galaxies found in 
\citet{Gadotti09}. 
The finding that all components have similar $g-i$ colours can be understood in
terms of the following properties:
($i$) the mean ages of disc and spheroid stars
differ typically by only $\sim 3$ Gyr (typical ages for the disc and spheroidal components are
$7.5-9$ and $11-12$ Gyr, respectively -- see table~2 of S09; with two 
two exceptions, Aq-A-5 and Aq-B-5, which have
very young discs), ($ii$) 
the distributions of  [Fe/H] of bulge and disc stars peak at low values, 
about $-0.5$ and $-1$, respectively, and differ by only a few tenths of a dex, 
the bulge stars 
generally having higher values of log [Fe/H], 
and ($iii$) the amount of leftover gas at $z=0$ is small, and
so is the amount of dust. As a result, the $g-i$ colours
reflect those of a stellar population which is blue due to low metallicity.
Overall, our results are  consistent with 
these simulated galaxies being disc-dominated, 
late-type systems, with low [Fe/H].

We have also compared the typical spatial sizes of discs, obtained
with the two decompositions. As explained above, we have converted
the half-mass radii of S09 into a corresponding scale-length
assuming an exponential surface density profile. 
The two estimates [columns (9) and (10) of Table~\ref{decomp_results},
respectively] agree well with the range observed in real
(late-type) galaxies (between $\sim 1$ and $\sim 8$ kpc -- 
see figure 18 in \citep{Gadotti09}), although they are in the high-end
tail. In real galaxies, the largest values for the disc scale-lengths are usually found
in galaxies with low B/T.

In columns (11)-(13) of Table~\ref{decomp_results}, we show the
half-mass radii of the spheroidal components
obtained with the KD, and the bulge effective
radii and S\'ersic indices from the PD, respectively. 
We discuss these results separately from those
presented before, noting that they may
be affected by numerical resolution as the
bulge scale-lengths obtained with both decompositions
are between $2$ and $4$ times the corresponding gravitational
softening (except for Aq-G-5, where the bulge effective
radius is of the order of the softening length).
 However, we do not find a correlation
between the gravitational softening length and the bulge
scale-lengths, so our results are not
necessarily softening dominated. 
The PD estimates for the bulge effective radii and 
the KD estimates for the spheroidal half-mass radii are 
similar, and are significantly larger than the bulge 
effective radii observed in real galaxies, which are 
typically about 1 kpc (see table 3 in \citealp{Gadotti09}).
As for the the S\'ersic indices, we find that $n\lesssim 1$ in {\em all} cases
(Table~\ref{decomp_results}). In  this aspect,
these bulges resemble pseudo-bulges rather than classical ones
(see, however, section~$4.2$ in \citealp{Gadotti09}).
This result is intriguing since these galaxies have undergone mergers
(although mainly minor ones), 
which are generally expected to produce classical bulges. The origin
of bulges in these simulations and its relation to
 S\'ersic indices will be investigated in detail in
a separate work.

\section{Conclusions}
\label{conclu}

We have analysed the $z=0$ outputs of eight cosmological,
hydrodynamical simulations of the formation of Milky Way-mass galaxies,
focusing on the study of the relative importance of the
main stellar components, i.e., bulges, discs and bars.
We have used two different analysis techniques:
a kinematic decomposition of stars into disc and spheroidal
components, widely used in simulation studies, 
and a photometric decomposition into bulge, disc and bar
components, as applied to observations.
Using both approaches, we estimated disc-to-total ratios
and disc and bulge scale-lengths. The photometric decomposition
also allowed us to compute the $g-i$ colours of the different components,
the bulge S\'ersic index, and bulge- and bar-to-total ratios, and to compare
them directly, and in a meaningful manner, to observational results obtained using the same techniques.

We found that the photometric and kinematic decompositions
predict different D/T ratios; those obtained with the former
method are systematically higher than those from the latter. 
The discrepancy cannot be attributed to the fact that contributions
to the different components are mass-weighted for the KD and
luminosity-weighted for the PD, and are not particular to
the band used in the PD (similar results are obtained for
the $i$ and $g$ bands).
In part, the discrepancy  can be explained considering that, in the KD,
disc stars are not present within the inner $\sim 2$ kpc, while
the PD assumes an exponential profile starting at
the centre of the galaxy.
This comparison is of relevance, since it 
indicates that the kinematic structure of real
galaxies may not be correctly inferred from the D/T ratios provided
by photometric analysis. When simulations are analysed in the same
way as observations, their properties appear closer to those of real
galaxies than when a KD is performed, although none of our simulations
comes close to reproducing a late-type, fully disc-dominated galaxy.

According to the PD,
half of the simulated galaxies have significant
bar components. A consequence of this result is 
that, even in galaxies with relatively low D/T ratios, 
bulges do not necessarily dominate over discs.
The presence of bars makes the comparison between the
two methods even harder since, in the kinematic approach,
bars are counted as part of the spheroidal component.
Results of the PD also show that
all components have similar $g-i$ colours, in the blue tail
of the observed distribution. This is understood in terms
of all components having low metallicity, relatively
similar ages, and low gas/dust content at $z=0$.

We  find  good agreement between the disc scale-lengths
obtained with our two methods. In the case of the bulges,
however, a conclusion is harder to make, since the
bulge scale-lengths are typically a few times the size
of the gravitational softening and thus may
be affected by resolution, although
there is no correlation between
bulge size and assumed softening.
We also find that S\'ersic indices are in all cases near unity,
 making the simulated
bulges more similar to observed pseudo-bulges
rather than to classical ones.

The results of this work suggest that, in order to be
meaningfully confronted with observations, results
from simulations need to be analysed following
observational techniques. On the other hand, simulations
contain valuable information that can be used to better
interpret observations, and they allow the physical processes
contributing to shape a galaxy's morphology to be studied.
We hope our work is a  step towards encouraging the exchange
of expertise between observers and simulators, since
a lot of physical insight can be gained from such an exchange.

\section*{Acknowledgments}

We thank the referee  for a thorough reading of this work and for helpful
comments and suggestions.
The authours acknowledge useful discussions with E. Athanassoula, P. Coelho,
R. de Jong, P. Ocvirk  and V. Springel.

\bibliographystyle{mn2emod}

\bibliography{biblio}

\end{document}